\documentclass[conference]{IEEEtran}
\IEEEoverridecommandlockouts

\usepackage{cite}
\usepackage{amsmath,amssymb,amsfonts}
\usepackage{algorithmic}
\usepackage{graphicx}
\usepackage{textcomp}
\usepackage{xcolor}

\usepackage{url}            
\usepackage{booktabs}       
\usepackage{amssymb} 
\usepackage{multicol}
\usepackage{multirow}
\usepackage{pifont}

\def\BibTeX{{\rm B\kern-.05em{\sc i\kern-.025em b}\kern-.08em
    T\kern-.1667em\lower.7ex\hbox{E}\kern-.125emX}}
\begin{document}

\title{Pan-protein Design Learning Enables Task-adaptive Generalization for Low-resource Enzyme Design
}

\author{\IEEEauthorblockN{1\textsuperscript{st} Jiangbin Zheng}
\IEEEauthorblockA{\textit{Zhejiang University} \\
\textit{Westlake University}\\
Hangzhou, China \\
zhengjiangbin@westlake.edu.cn}
\and
\IEEEauthorblockN{2\textsuperscript{nd} Ge Wang}
\IEEEauthorblockA{\textit{Zhejiang University} \\
\textit{Westlake University} \\
Hangzhou, China \\
wangge@westlake.edu.cn}
\and
\IEEEauthorblockN{3\textsuperscript{rd} Han Zhang}
\IEEEauthorblockA{\textit{Westlake University} \\
Hangzhou, China \\
zhanghan36@westlake.edu.cn}
\and
\and
\IEEEauthorblockN{4\textsuperscript{th} Stan Z. Li*}
\IEEEauthorblockA{\textit{Westlake University} \\
Hangzhou, China \\
stan.zq.li@westlake.edu.cn\\
*Corresponding Author
}
}

\maketitle

\begin{abstract}
Computational protein design (CPD) offers transformative potential for bioengineering, but current deep CPD models, focused on universal domains, struggle with function-specific designs. This work introduces a novel CPD paradigm tailored for functional design tasks, particularly for enzymes-a key protein class often lacking specific application efficiency. To address structural data scarcity, we present CrossDesign, a domain-adaptive framework that leverages pretrained protein language models (PPLMs). By aligning protein structures with sequences, CrossDesign transfers pretrained knowledge to structure models, overcoming the limitations of limited structural data. The framework combines autoregressive (AR) and non-autoregressive (NAR) states in its encoder-decoder architecture, applying it to enzyme datasets and pan-proteins. Experimental results highlight CrossDesign's superior performance and robustness, especially with out-of-domain enzymes. Additionally, the model excels in fitness prediction when tested on large-scale mutation data, showcasing its stability.
\end{abstract}


\begin{figure*}[t!]
 \centering
 \includegraphics[width=0.82\linewidth]{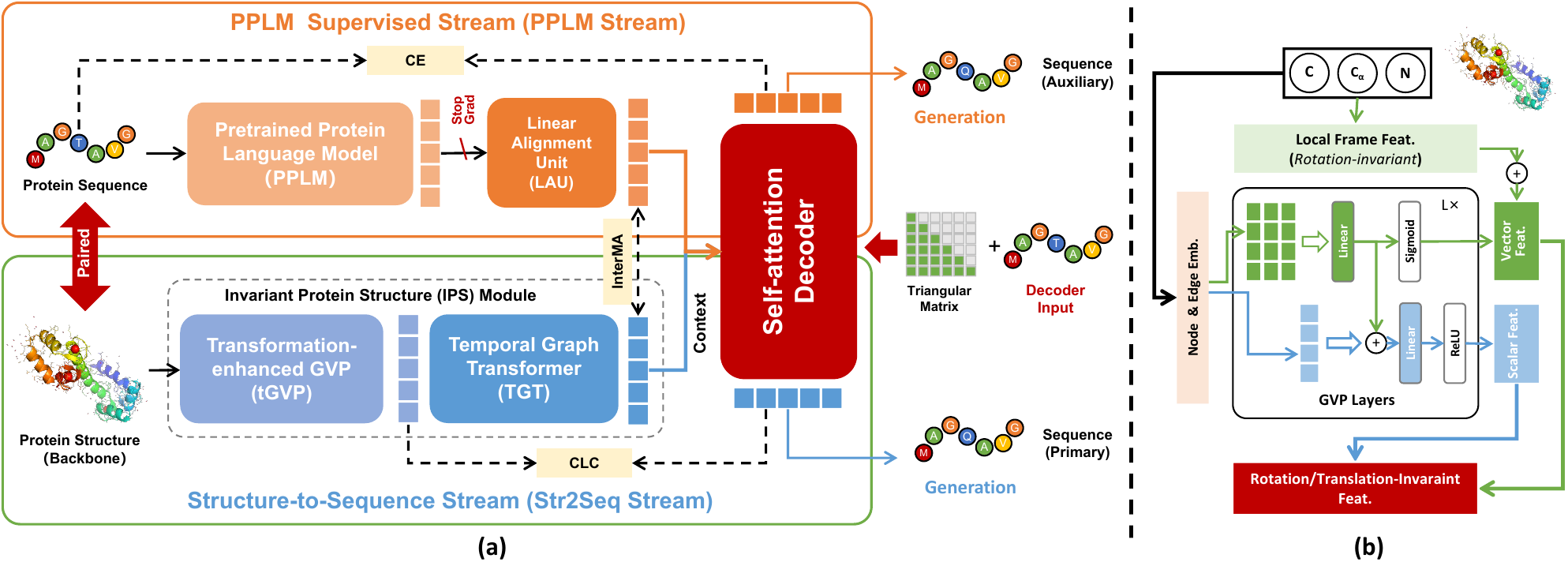}
 \vspace{-1em}
 \caption{(a). The proposed CrossDesign framework contains a Structure-to-Sequence Stream (Str2Seq Stream) and an auxiliary PPLM Supervised Stream (PPLM Stream). The PPLM stream is used to augment the Str2Seq stream using prior language knowledge, while the two streams share a decoder. CLC: Cross-Layer Consistency; InterMA: inter cross-modal alignment. \textbf{Note that} once the training is done, the framework performs the Str2Seq stream only for sampling. (b) Schematic diagram of Transformation-enhanced GVP (tGVP).}
 \label{fig:1}
 \vspace{-1em}
\end{figure*}

\section{Introduction}
\label{sec:intro}
Computational protein design (CPD) aims to generate amino acid sequences given protein backbones,and plays a pivotal role in bioengineering, particularly in drug discovery and enzyme design.
While deep CPD models have ventured into de novo protein design, they remain confined to the general pan-protein domain.
In response, this work proposes a novel CPD paradigm explicitly designed for functional tasks, emphasizing enzymes-a crucial protein class frequently underserving specific application requirements. 
One of primary tasks in enzyme engineering is to improve property-specific enzymes by mutations or redesign.
To achieve this, a CPD model must possess strong representation and generalization and offer a flexible architecture for expanded downstream enzyme tasks.

Data scarcity, particularly the lack of sizable structure-sequence datasets, poses a significant bottleneck, leading to issues like undertraining or overfitting. Despite efforts to enhance equivariance and invariance in structural modules, practical protein design models with existing low-resource data remain elusive.
This work considers transfer learning as a potential solution, with a focus on the transfer of pretrained structure models. However, the scarcity of such models with powerful representation capability limits their use. Redirecting attention to pretrained protein language models (PPLMs), which have achieved success in representing protein sequences, provides an alternative. PPLMs, trained on extensive sequence datasets \cite{rao2020transformer,rives2021biological,jumper2021highly,lin2022language}, offer valuable implicit structural information for CPD, although no attempts have been made so far to the best of our knowledge.
Nevertheless, CPD is essentially a cross-modal structure-to-sequence task, but PPLMs are difficult to be applied directly since they represent the sequence modality only.
Fortunately, emerging text-visual cross-modal modeling techniques such as CLIP \cite{radford2021learning,ramesh2022hierarchical} have opened a promising path, and robust protein structure modeling is inspired to incorporate valuable PPLM prior for enhancement.

To overcome data scarcity and cross-domain requirement, we propose a novel transformation-enhanced domain-adaptive protein design framework, \textit{CrossDesign}, incorporating cross-modal alignment between PPLM and protein structures, as shown in Figure~\ref{fig:1}. Additionally, we implement inter-modality alignment (InterMA) to align these modalities and apply cross-layer consistency (CLC) to regularize the process.  
While fully utilizing PPLM’s supervised signal during training, our model eliminates the need for PPLM during inference, significantly reducing computational costs.
The encoder-decoder framework facilitates easy extension of the sampling regime. For conditional CPD tasks with unknown sequences, autoregressive (AR) sampling is used, leveraging the lower triangle mask during training. For proteins with known sequences—crucial for downstream function prediction, such as mutation effects, non-autoregressive (NAR) parallel computing is applied without additional training.

The main contributions are as follows:

$\bullet$ The proposed CrossDesign innovatively incorporates PPLMs with cross-modal alignment and cross-layer constraint in protein design, highlighting their potential in enhancing protein structural representation.
 
$\bullet$  The skillfully crafted architectural design is well-suited for cross-domain CPD and mutation tasks. 

$\bullet$ Robust performances across diverse in-domain and out-of-domain tasks demonstrate versatility and effectiveness.

$\bullet$ The collection of enzyme datasets and benchmarks serves to narrow the gap between generic proteins and specific functional proteins for structure-sequence transformation.

\section{Methods}
\label{sec:methods}
As shown in Figure~\ref{fig:1}(a), CrossDesign is composed of the Structure-to-Sequence Stream (Str2Seq stream), and the auxiliary PPLM Sequence Stream (PPLM stream).
To bridge these two independent streams, a linear alignment unit (LAU) and a shared decoder are utilized.
CrossDesign is trained in an end-to-end manner. The inference pipeline relies only on the Str2Seq stream and does not require auxiliary PPLM.

Formally, $X=\{x_1, x_2, \cdots, x_n \}$ denote the protein atomic coordinates given a structure-sequence pair, where $n$ is the sequence length, and $x_i \in \mathbb{R}^{3 \times 3}$ represents the $i-$th coordinates of the amino acid residue, consisting of N, C$_{\alpha}$, and C atoms. 
The corresponding primary generated sequence in Str2Seq stream is represented as $Y_p = \{y_1, y_2, \cdots, y_n \} \in \mathbb{R}^{n}$, where $Y_i$ denotes the $i-$th amino acid residue in textual form.
While the corresponding auxiliary generated sequence in PPLM stream is represented as $Y_a = \{y’_1, y’_2, \cdots, y’_n \} \in \mathbb{R}^{n}$.
The reference native sequence of $Y$ is $\hat{Y} = \{ \hat{y}_1, \hat{y}_2, \cdots, \hat{y}_n \} \in \mathbb{R}^{n}$. 
$X$ and $\hat{Y}$ is a ground-truth structure-sequence pair.

\subsubsection{Main Structure-to-sequence Stream}
The Str2Seq stream is a completed and independent protein design network with an Invariant Protein Structure (IPS) module and a decoder, as shown in Figure~\ref{fig:1}(a). This stream aims to generate sequences given backbone coordinates.
The geometric Transformation-enhanced GVP (tGVP) is a key component of IPS. Among geometric models, the lightweight GVP~\cite{jing2020learning} is chosen with properties of equivariance and invariance for rigid bodies.
Temporal GraphTransformer (TGT) consists of a graph convolution network and a generic Transformer encoder to enhance the temporal interactions while aligning Str2Seq and PPLM streams and act as a linker between the tGVP and the decoder in Str2Seq stream. 
The self-attention decoder is a generic Transformer decoder with triangular masks \cite{vaswani2017attention}. The only modification is the use of learnable positional embeddings instead of sinusoidal positional embeddings. The decoder module is shared between two streams.
The outputs of TGT are fed into the decoder, also used in InterMA.

\subsubsection{Auxiliary PPLM Supervised Stream}
The PPLM stream is an auxiliary network for augmenting Str2Seq stream. The entire PPLM stream is essentially an inner-loop asymmetric auto-encoder (asyAE) model.
The PPLM is the core component for introducing prior language knowledge, which is fixed and untrainable. We transfer the off-the-shelf PPLM trained on large-scale protein sequence datasets.
The LAU module is an adapter simply implemented as a fully connected MLP with two hidden layers. The LAU forces the conversion of the dimensions of PPLM outputs to align with TGT outputs. Also, LAU serves the InterMA loss, acting as an intermediary between the Str2Seq and PPLM streams and assuming the role of semantic supervision. 
The decoder is an auxiliary bridge in the PPLM stream, shared with Str2Seq stream, which takes LAU's outputs as inputs and generates auxiliary protein sequences based on asyAE reconstruction loss.


\subsubsection{Primary Generation Objective}
In Str2Seq stream, the protein backbone $X$ is first input to the tGVP module to obtain the geometric structural features $Z_\text{struc}$ with a hidden dimension $d$.
Then the structural features $Z_\text{struc}$ are fed to the TGT blocks and decoder sequentially through Str2Seq stream. 
The features obtained by the self-attention TGT blocks are referred to as alignment features $Z_\text{align}$. The features obtained by the decoder are denoted as contextual features $Z_\text{context}$. Finally, the following MLP layers generate the output distribution $D_{\text{logits-p}}$ of the primary generated sequence ($M$=20 denotes 20 types of amino acids) as:
\begin{equation}
\left \{
    \begin{array}{rl}
         Z_\text{struc} &  = \text{tGVP}(X) \in \mathbb{R}^{n \times d},\\
    Z_\text{align} &  = \text{TGT}(Z_\text{struc}) \in \mathbb{R}^{n \times d},\\
    Z_\text{context} &  = \text{Decoder}(Z_\text{align}) \in \mathbb{R}^{n \times d},\\
     D_{\text{logits-p}} & = \text{MLP}(Z_\text{context}) \in \mathbb{R}^{n \times M},\\
     p(Y_p|X) & = \text{argmax}\big( \text{Softmax}(D_{\text{logits-p}})\big) \in \mathbb{R}^{n}.
    \end{array}
\right.
\end{equation}

The cross-entropy (CE) loss is commonly used to measure the discrepancy between two probability distributions. We observed that emphasizing the salient features of the output distribution can further improve the performances. To achieve this, an exponential cross entropy (expCE) loss is used to accentuate the spiking effect of the output distribution as:
\begin{equation}
\mathcal{L}_\text{expCE} = \sum_{b=0}^{B}\exp\left(\text{CE}(\text{Softmax}(D_{\text{logits-p}}^{(b)}), \hat{Y}^{(b)})\right),
\end{equation}
where $B$ denotes the number of elements in a batch.

\subsubsection{Cross-layer and Cross-modal Objective}
Multi-modal studies have shown that it makes sense to explicitly enforce consistency between different modalities~\cite{zheng2023cvt,zheng2024metaenzyme,zheng2024progressive,zheng2023mmdesign,zheng2023lightweight,zheng2022using,zheng2021enhancing,zheng2020improved,zheng2022leveraging,kamal2019technical,tong2020document,xia2024understanding,huang2024protein,hu2023learning,hu2022protein,huang2023data,xia2024discognn,wu2020fuzzy,zuo2022c2slr,hao2021self}. 
In the Str2Seq stream, structural features $Z_\text{struc}$ from the tGVP module and the contextual features $Z_\text{context}$ from the decoder are considered as two different internal modalities. As shown in Figure~\ref{fig:1}, to enforce alignment constraints across modalities, we propose a cross-layer CLC loss, which is implemented as a knowledge distillation (KL-divergence here), where the entire  $Z_\text{context}$ and $Z_\text{struc}$ are treated as teacher and student models. A high temperature $\tau=8$ is adopted to `soften' the probability distribution from potential spike responses\cite{zheng2022using}. The distillation process of cross-layer CLC is expressed as:
\begin{equation}
\mathcal{L}_\text{CLC} = \text{KL}(\text{Softmax}(\frac{Z_\text{struc}}{\tau}), \text{Softmax}(\frac{Z_\text{context}}{\tau})).
\end{equation}

Similarly, we employ knowledge distillation for InterMA loss to align Str2Seq and PPLM streams.
In PPLM stream, the native sequence $\hat{Y}$ is fed to the PPLM, LAU, and decoder sequentially. 
The LAU output is denoted as another type of context $Z'_\text{context}$.
Unlike CLC, we use geometric-temporal features ($Z_\text{align}$) in contextual modality instead of geometric features ($Z_\text{struc}$) of the tGVP, since CLC is a cross-stream cross-modal alignment. The alignment of InterMA as:
\begin{equation}
\begin{split}
   Z'_\text{context} &= \text{LAU}(\text{PPLM}(\hat{Y})) \in \mathbb{R}^{n \times d},\\
    \mathcal{L}_\text{InterMA} &= \text{KL}(\text{Softmax}(\frac{Z_\text{align}}{\tau}), \text{Softmax}(\frac{Z'_\text{context}}{\tau})). 
\end{split}
\end{equation}

\subsubsection{Inner-loop Sequence Reconstruction}
To enable the trainable modules (LAU and Decoder) in PPLM stream to fit faster and learn potential contextual knowledge, we employ a Seq2Seq recovery mode as a pseudo-machine translation process since PPLM stream can be considered an asyAE model, as shown in Figure~\ref{fig:1}.
To reconstruct auxiliary sequences $Y_a$ in PPLM stream, the contextual features $Z'_\text{context}$ from the LAU are fed to the shared decoder. And the $Y_a$ corresponds to the generated probability $D_\text{logits-a}$. The final sequence reconstruction loss is denoted as:
\begin{equation}
\label{eq:aeloss}
\mathcal{L}_{\textrm{asyAELoss}} = \textrm{CE} \big(\textrm{Softmax} (D_\textrm{logits-a}), \hat{Y} \big).
\end{equation}

\subsection{Sampling Regimes for Cross-domain Tasks}

Different sampling regimes vary from cross-domain tasks based on AR and NAR modes.

\subsubsection{AR Decoding for CPD}
Given a candidate structure $X_{c}$ for generating the unknown sequence $Y_{sample}$ as:
\begin{equation}
    p(Y_{sample}|X_{c}) = \prod \limits_{i=0}^n p(y_i | y_{i-1}, \cdots, y_0; X_{c}).
\end{equation}
This density is modeled autoregressive through the Vanilla Transformer decoder paradigm. Using AR decoding we can typically conduct CPD tasks for general proteins and function-specific enzymes. 

\subsubsection{NAR Decoding for Fitness}
Compared with AR decoding, NAR is applicable for mutation effect utilizing zero-shot learning. Given a known enzyme backbone $X_{wild}$ and the corresponding mutant sequence $Y_{mut}$, the mutant correlation can be scored based on marginal probability $Prob_{mut}$ as:
\begin{equation}
    Prob_{mut} = p(\text{logits}|\text{Decoder}(Y_{mut});\text{IPS}(X_{wild})).
\end{equation}
Contrary to AR decoding, NAR can perform inference tasks faster with parallel computing.

\section{Experiments}
\subsection{Settings}

\textbf{Generic Protein Datasets.} 1) The \textit{CATH} dataset \cite{ingraham2019generative} includes training, validation, and testing splits with 18204, 608, and 1120 structure-sequence pairs, respectively. CATH specifically refers to CATH 4.2 if not otherwise specified. 2) We also report results on \textit{Ts50} \& \textit{Ts500} \cite{li2014direct} to assess the generalization. No canonical training and validation sets exist for Ts50 or Ts500.

\textbf{Functional Enzyme Datasets.}
1) \textit{EnzPetDB}: We collected 212 proteins described to breakdown plastics from PlasticDB.
2) \textit{EnzFoldDB}: We consolidated 691 experimentally-determined structure-sequence paired enzymes with 10 folds from RCSB. Each fold has an equal number of enzymes.
3) \textit{ProteinGym}~\cite{notin2023proteingym} is specifically designed for protein fitness prediction and design, which encompasses a broad collection of over 178 standardized deep mutational scanning assays, spanning millions of mutated sequences.

\subsection{In-domain Evaluation on Generic Proteins.}
Overall, CrossDesign outperforms baselines consistently across all test sets with balanced performances, showing strong generalization and robustness. In addition, a pretrained strategy based on prior knowledge transfer with a small training set can outperform the strategy based on large data training such as ESM-IF (Group 2 vs. Group 3).

\textbf{Performances on comprehensive protein data.}
Table~\ref{tab:1} reports the perplexity (lower is better) and recovery (higher is better) scores on the generic dataset CATH test set (``ALL''). CrossDesign achieves the best performance with a perplexity of 3.67 and a recovery of 57.11\%, outperforming consistently the state-of-the-art ProteinMPNN, PiFold (Group 1), and ESM-IF (Group 2) by a significant margin.

\textbf{Performances vary from the types of chains.}
Following \cite{ingraham2019generative}, we evaluate the models on the subsets of the CATH test set, i.e., ``Short'' dataset (protein sequence length $\leq$ 100 residues) and ``Single-chain'' dataset (single-chain proteins recorded in the Protein Data Bank) as shown in Table~\ref{tab:1}. CrossDesign consistently exceeds baselines on both ``Single-chain'' and ``Short'' datasets. In particular, in terms of perplexity, CrossDesign (Short: 5.17; Single-chain: 5.31) significantly outperforms PiFold (Short: 6.04; Single-chain: 6.31) and ProteinMPNN (Short: 6.21; Single-chain: 6.68).

\textbf{Domain-unknown Generalization.}
To fully compare the generalizability of the different models on the out-of-domain datasets, we report the results on the Ts50 and Ts500 datasets in Table~\ref{tab:1} since they are completely different from the training data. CrossDesign still achieves consistent improvements for Ts50 and Ts500, where the recovery score in Ts50 breaks 60\% for the first time, and the score in Ts500 is close to 61\%.

\begin{table}[t!]
\centering
\scalebox{0.55}{
\begin{tabular}{clcccccccccc} 
 \toprule

 \multirow{2}{*}{\textbf{\#}} & \multirow{2}{*}{\textbf{Models}} & \multicolumn{5}{c}{\textbf{Perplexity}} & \multicolumn{5}{c}{\textbf{AAR(\%)}} \\ 

 \cmidrule(lr){3-7}\cmidrule(lr){8-12}
 
 & & \textbf{All} & \textbf{Short} & \textbf{Single-chain} & \textbf{Ts50} & \textbf{Ts500} & \textbf{All}& \textbf{Short} & \textbf{Single-chain} & \textbf{Ts50} & \textbf{Ts500} \\ 
 \midrule
 
 \multirow{12}{*}{1} 
 & Natural frequencies \cite{hsu2022learning} & 17.97 & 18.12 & 18.03 & - & - & 9.5 & 9.6 & 9.0 & - & -\\
 & SPIN \cite{li2014direct} & - & - & - & - & - & - & - & - & 30.3 & 30.3 \\
 & SPIN2 \cite{o2018spin2} & - & 12.11 & 12.61 & - & - & - & - &- & 33.6 &36.6 \\
 & StrucTransformer \cite{ingraham2019generative} & 6.85 & 8.54 & 9.03 & 5.60 & 5.16 & 36.4 & 28.3 & 27.6 & 42.40 &44.66 \\
 & Structured GNN \cite{jing2020learning} & 6.55 & 8.31	& 8.88 & 5.40 & 4.98 & 37.3 & 28.4 & 28.1 & 43.89 & 45.69\\
 
 & ESM-IF \cite{hsu2022learning} & 6.44 & 8.18 & 6.33 & - & - & 38.3 & 31.3 & 38.5 & - &-\\
 
 & GCV \cite{tan2022generative} & 6.05 & 7.09 & 7.49 & 5.09 & 4.72 & 37.64 & 32.62 & 31.10 & 47.02 & 47.74\\

 & GVP-GNN-large \cite{jing2020learning} & 6.17 & 7.68	 & \underline{6.12} & - & - & 39.2 & 32.6 & \underline{39.4} & - &-\\
 & GVP-GNN \cite{jing2020learning} & 5.29 & 7.10 & 7.44 & 4.71 & 4.20 & 40.2 & 32.1 & 32.0 & 44.14 & 49.14 \\
 
 & AlphaDesign \cite{gao2022alphadesign} & 6.30 & 7.32 & 7.63 & 5.25 & 4.93 & 41.31 & 34.16 & 32.66 & 48.36 & 49.23\\
 & ProteinMPNN \cite{dauparas2022robust} & 4.61 & 6.21	 & 6.68 & 3.93 & 3.53 & 45.96 & 36.35 & 34.43 & 54.43 & 58.08\\
 
 & PiFold \cite{gao2022pifold} & 4.55 & \underline{6.04} & 6.31 & 3.86 & 3.44 & 51.66 & \underline{39.84} & 38.53 & \underline{58.72} & 60.42\\
 \midrule
 
 \multirow{1}{*}{2}
 & ESM-IF \cite{hsu2022learning} & \underline{3.99} & 6.30	 & 6.29 & \underline{3.43} & \underline{3.34} & \underline{52.51} & 34.74 & 34.25 & 56.66 & \underline{60.85}\\
 \midrule
 
 \multirow{1}{*}{3}
 
 & \textbf{Ours (CrossDesign)} & \textbf{3.67} & \textbf{5.17} & \textbf{5.31} & \textbf{2.99} & \textbf{3.24} & \textbf{57.11} & \textbf{40.71} & \textbf{39.47} & \textbf{60.04} & \textbf{60.90}\\
	 \bottomrule
 
\end{tabular}
}
\caption{Comparison on the CATH, Ts50, and Ts500 datasets. The best results are \textbf{bolded}, and the second best \underline{underlined}. Group 1 \& 3 are trained on a tiny CATH training set only, while Group 2 is trained on a large CATH+AlaphaDB training set. The \textit{Short} and \textit{Single-chain} are subsets of CATH test set \textit{ALL}.}
\vspace{-2.0em}
\label{tab:1}
\end{table}

\begin{figure}
\centering
    \centering
    \includegraphics[width=0.78\linewidth]{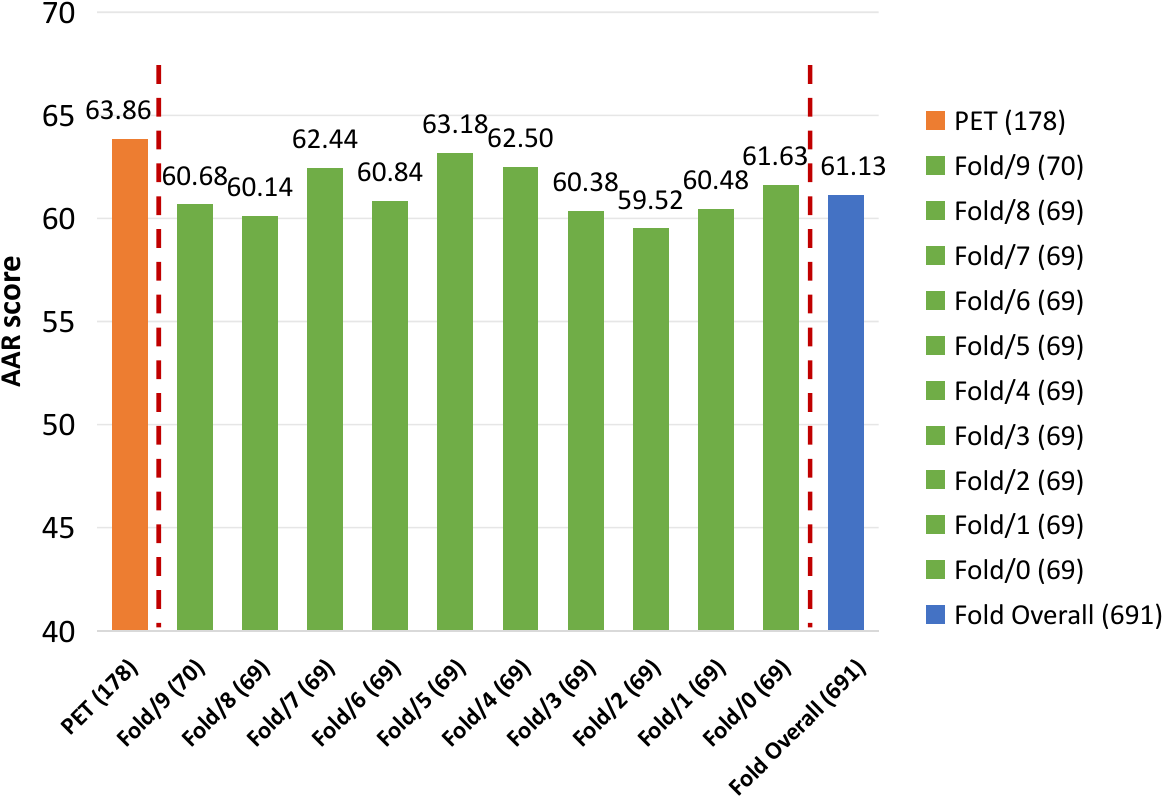}
    \vspace{-1em}
    \caption{AAR of enzyme design. Orange for EnzPetDB; Green and blue for EnzFoldDB. \textit{EC/i(N)} denotes the evaluation in the i-th fold of the number N.}
    \vspace{-1em}
    \label{fig:arr_enzymes}
\end{figure}

\subsection{Out-of-Domain Enzyme Design}

\subsubsection{Conditional CPD for PET Enzymes}

Polyethylene terephthalate (PET) is a synthetic plastic polymer widely used globally. Due to its ester bonds and aromatic nuclei, PET is chemically inert and highly resistant to degradation, leading to environmental and health concerns. Plastic waste poses a significant ecological challenge, and enzymatic degradation offers a promising, eco-friendly method for recycling polyester waste. However, knowledge of PET-degrading enzymes remains limited. Machine learning-assisted methods could accelerate the discovery of PET hydrolases \cite{lu2022machine}. To support this effort, we curated a comprehensive dataset for analyzing PET hydrolases using CPD, which will serve as a valuable resource for future research. As shown in the orange bar of Figure~\ref{fig:arr_enzymes}, CPD analysis achieved AAR scores up to 64.86\% higher than those in in-domain settings, indicating strong generalization for unseen PET datasets and paving the way for further data analysis and enzyme modifications.

\subsubsection{Conditional CPD for Fold-aware Enzymes}
Based on enzyme categorization criteria, enzymes are classifiable into different groups based on their fold types. To assess the performance of enzymes at various fold levels in the CPD task, we categorized them into 10 distinct classes as illustrated in Fig.~\ref{fig:arr_enzymes}. The green bars represent fold-specific evaluations, with AAR scores for each fold consistently hovering around 60\%, ranging from 59.52\% (Fold/2) to 63.18\% (Fold/5). Notably, these scores surpass the level of out-of-domain generalization observed in the generic protein dataset.
In contrast, the blue bar signifies the fold-agnostic display, amalgamating all folds and providing an average assessment of 61.15\%. This result underscores a consistently high overall performance level across diverse fold types.

\subsubsection{Zero-shot Learning for Mutation Effect Prediction}\label{subsec:mutant}

\textbf{Mutant Scoring with NAR decoding.}
Contrary to language models~\cite{meier2021language}, our model scores mutation effects using a different method since the special encoder-decoder architecture. Given the wildtype enzyme structure feature $Z_{wild}$ and the mutant sequence $Y_{mut}$ corresponding to the wildtype sequence $Y_{wild}$, the NAR decoding generate mutant marginal probability $Prob_{mut}$ as:
\begin{equation}
\begin{split}
    Prob_{mut} & = p(\text{logits}_{mut}|Z_{wild}; Y_{mut}),\\
    \text{Effect}_{mut} &= \text{CE}(Prob_{mut}, Y_{wild}),
\end{split}    
\end{equation}
This scoring method can compute either single or multiple mutations. CE is used to be consistent with the training setup, which is used to assess the difference between mutant distribution and wildtype sequence.

\textbf{Performances on ProteinGym}.
To ensure a fair comparison, we selected prominent protein language models and inverse folding models as baselines, consistent with ProteinGym~\cite{notin2023proteingym}. All baselines are computed using zero-shot learning for mutation effects. Assessing all 217 ProteinGym proteins containing millions of mutations, our model consistently achieved the best matching rank (0.445). In comparison, the optimal protein language model (VESPA) averaged $\rho$ of 0.437, and the optimal inverse folding model (ESM-IF) scored 0.422. This performance edge may be attributed to our model's dual advantage, encompassing both language modeling and structural modeling within the proposed paradigm.
To further validate performance \textit{across functions and taxa}, we compared CrossDesign with other inverse folding models. In results categorized by function, all models exhibit superior performance in terms of Stability. Notably, our model outperforms others, particularly in Binding (0.412) and Stability (0.636), albeit with slightly lower scores in other aspects. In results categorized by taxon, our model surpasses ESM-IF for Prokaryotic Viruses and slightly lags behind ESM-IF in other categories. Overall, the results are consistently excellent.

\section{Conclusions and Limitations}
This work contributes to advancing CPD methodologies, bridging the gap between universal protein design and the nuanced requirements of specific functional protein engineering.
While CrossDesign exhibits versatility with AR and NAR decoding modes, enhancing capabilities for functional protein prediction, certain \textbf{limitations} persist:
1) Limited functional protein datasets were utilized, due to the challenges in collecting enzyme datasets. 2) Augmenting the structure-sequence training data holds the potential to express more robust architectures. 3) The verification of mutation effects lacks wet experiment-assisted validation.



\bibliographystyle{plain}
\bibliography{my_citation,self_citation}

\end{document}